\documentclass[aps,pre,groupedaddress,showpacs]{revtex4}
\usepackage{graphicx}% Include figure files
\usepackage{dcolumn}% Align table columns on decimal point
\usepackage{bm}% bold math
\usepackage{amssymb}
\usepackage{amsmath}

\begin{document}

\title{Comprehensive study of phase transitions in relaxational systems with
field-dependent coefficients}
\author{J. Buceta and Katja Lindenberg}
\affiliation{Department of Chemistry and Biochemistry, and Institute for Nonlinear Science,
University of California San Diego, 9500 Gilman Drive, La Jolla, CA 92093-0340, USA}

\begin{abstract}
We present a comprehensive study of phase transitions in
single-field systems that relax to a non-equilibrium global
steady state.  The mechanism we focus on 
is not the so-called Stratonovich drift combined with collective effects,
but is instead similar to the one
associated with noise-induced transitions \emph{a la} Horsthemke-Lefever in
zero-dimensional systems. As a consequence, the noise interpretation
(e.g., It\^{o} vs Stratonvich) merely shifts the phase boundaries.
With the help of a mean-field approximation, we present a broad
qualitative picture of the various phase diagrams that can be found in
these systems.  To complement the theoretical analysis we
present numerical simulations that confirm the findings of the
mean-field theory. 
\end{abstract}


\maketitle

\section{Introduction}

Equilibrium and non-equilibrium relaxational dynamics play an important role
in many critical phenomena \cite{halperin}. Typically, relaxational flows
drive the evolution of the system to the equilibria
determined by a Lyapunov energy functional $\mathcal{F}$ that depends
on local
potential functionals and on the interactions in the
system~\cite{jordi, montagne}.  The celebrated $\phi^4$ functional is
a paradigmatic example of such a
potential functional and gives rise to well-known equilibrium models such as
the so-called model A (a coarse-grained version of the Ising model), 
and model B (an archetype of phase separation dynamics), among
others~\cite{halperin, jordi}.

Typical relaxational models describe the flow of
a field $\varphi_i(t)$ defined on a $d$-dimensional square lattice via a
Langevin equation of the form
\begin{equation}
\dot{\varphi}_i(t) =-\Gamma
\frac{\delta \mathcal{F}\left( \left\{ \varphi \right\} \right) }
{\delta \varphi_i(t)}+\Gamma^{1/2}\xi_i(t).
\label{phiold}
\end{equation}
Here $i$ labels a lattice site, $\Gamma$ is a positive constant, 
$(\left\{\varphi\right\})\equiv (\varphi_1,\dots,\varphi_N)$ denotes
the entire set of fields, and $\xi_i$ are Gaussian white noises with
zero mean and correlation functions
\begin{equation}
\langle \xi_i(t) \xi_j(t^{\prime})
\rangle =\sigma^2\delta_{ij}\delta(t-t^{\prime }).
\label{fdr}
\end{equation}
Typically, the functional $\mathcal{F}$ consists of a local potential 
$V(\varphi)$ and an interaction term,
\begin{equation}
\mathcal{F}\left(\left\{ \varphi \right\} \right) =\sum_i\left( V(\varphi_i)
+\frac{K}{8d}\sum_{\langle ij\rangle }(\varphi_{j}-\varphi_i)^2\right),
\label{F}
\end{equation}
where $K$ is the coupling coefficient. The left-most sum in
Eq.~(\ref{F}) runs over 
all lattice sites and the right-most sum over the $2d$
nearest neighbors of a given site $i$.
A keystone in this formalism is the
link (the ``fluctuation-dissipation relation'') between the intensity
$\Gamma\sigma^2$ of the fluctuating contribution and the relaxation
parameter $\Gamma$ through the constant $\sigma^2$ that in
equilibrium systems is
proportional to the temperature.  The relaxation coefficient $\Gamma$ 
affects how fast the system relaxes to the global steady state.  

Recent studies have revealed the importance of field-dependent
relaxation coefficients $\Gamma(\varphi)$ in the dynamics of
these systems~\cite{marta}.  A generic
description for such systems is provided by the Langevin equation
\begin{equation}
\dot{\varphi}_i(t)=-\Gamma\left( \varphi_{i}(t) \right)
\frac{\delta \mathcal{F}\left( \left\{\varphi\right\}\right) }
{\delta \varphi_i(t) }+\left[ \Gamma \left( \varphi_i(t)\right)
\right] ^{\frac{1}{2}}\xi_i(t),
\label{phi}
\end{equation}
where $\Gamma(\varphi)$ and $[\Gamma(\varphi)]^{1/2}$ are both positive.
For example, it has been shown that
relaxational flows driven by field-dependent coefficients may
present \emph{inverted phase diagrams} where ordering effects increase with
the intensity of the fluctuations. This behavior has been observed in
polymer mixtures where spinodal decomposition, i.e., phase separation,
increases with increasing temperature~\cite{snyder}.

The importance of these flows 
may be even more pronounced in nonequilibrium systems, and
goes well beyond the scenarios that lead to inverted phase diagrams.
One example is that of pure noise-induced phase
transitions~\cite{marta}.  Such phase transitions exhibit the
striking feature that noise is the \emph{crucial} element
responsible for the appearance of ordered phases that 
disappear in the absence of noise.
The first mechanism identified in the literature leading
to such behavior~\cite{chris} relied on a combination of the
so-called Stratonovich drift that arises under this particular
interpretation of the noise, and collective effects. The
Stratonovich drift in these systems
leads to opposite dynamical responses at short and long
time scales.  Collective effects generated by
the coupling among the field elements can
amplify short time instabilities (that would die away in the absence of
coupling), thus leading to the observed noise-induced phase transitions.
As a consequence, there was originally a widespread belief 
that noise-induced phase transitions could only be found in systems
where there are no noise-induced transitions~\cite{horsthemke},
since the latter transitions occur in zero-dimensional
(uncoupled) systems. Recent studies involving relaxational flows with
field-dependent relaxation coefficients have shown otherwise by
presenting a system where both a transition (zero-dimensional)
and a phase-transition (coupled systems) are induced
by the same source of noise~\cite{marta}. In fact, the mechanism is not
attributable to the Stratonovich drift because these transitions occur
independently of the noise interpretation~\cite{carrillo}. Moreover, the
same mechanism has been extended to pattern formation
phenomena~\cite{buceta},
generalizing a previous mechanism based on the Stratonovich
drift~\cite{buceta2}.

A thorough understanding of relaxational models driven by
field-dependent coefficients is therefore important for a number of reasons.
They play a relevant role in critical phenomena, they
may explain situations where inverted phase diagrams are obtained, and
they constitute a generalization of the seminal work of Horsthemke
and Lefever on noise-induced transitions to noise-induced phase transitions
in coupled systems.  

Whereas specific non-equilibrium relaxational models driven by
field-dependent coefficients have been considered in the literature,
herein we present a more general analysis of such models, whereby the
specific cases considered earlier become part of a broad panorama.  We
ask two questions: (1) What are the circumstances (features of the
model, values of the control parameters) that lead to purely
noise-induced phase transitions? (2) What is the nature of the phases
that can occur in these systems, and what are the features of the model
that determine these phases?  In answering these questions, we discuss the
possible phase diagrams that can be obtained and show that their overall
structure depends on geometrical properties such
as the balance of convexities of the
local potentials and of the field-dependent coefficients. Furthermore, we
show how multistability can be induced by noise. 
Our point of departure for this analysis is the equation obtained by
implementing the functional derivative of $\mathcal{F}$ indicated in 
Eq.~\eqref{phi},
\begin{equation}
\dot{\varphi}_i(t)=\Gamma \left(\varphi_i\right)\left(
-\frac{\partial V\left(\varphi_i\right)}{\partial \varphi_i}
+\mathcal{L}\varphi_i\right)+\left[\Gamma \left(\varphi_i\right)
\right]^{\frac{1}{2}}\xi_i(t),
\label{phi2}
\end{equation}
$\mathcal{L}$ being the discrete version of the diffusion Laplacian
operator,
\begin{equation}
\mathcal{L}\varphi_i=\frac{K}{2d}\sum_{\langle ij\rangle}
\left(\varphi_j-\varphi_i\right) .
\end{equation}

The paper is organized as follows.
In Sec.~\ref{meanfield} 
we construct the mean field approximation to Eq.~\eqref{phi2} and
establish the different phases that may appear in the model. A discussion
of the possible phase boundaries between these phases, and
the order of the transitions, are presented in Sec.~\ref{boundaries}. The
structure of the resulting phase diagrams is presented in
Sec.~\ref{overall}, as is a specific illustration that corroborates our
more general analysis.  Sections ~\ref{boundaries} and \ref{overall} are
supplemented by an appendix where we show that a particular type of
transition, while it may occur, is necessarily an isolated point in the
phase diagram. In Sec.~\ref{numerical}, simulations of the full model
corroborate some of our most striking results. We conclude
in Sec.~\ref{conclusion} with a summary and some directions for
future research. 

\section{Phase Transitions: Mean-Field Analysis}
\label{meanfield}

We focus our analysis on systems that may undergo Ising-like phase
transitions.  A convenient order parameter to characterize the
phase transitions is akin to the \emph{magnetization}, 
\begin{equation}
m=\left| \left\langle \varphi \right\rangle \right| ,
\end{equation}
where the brackets indicate both a spatial and a temporal
average of the field in the steady state.  Ordered states are associated
with $m\neq 0$.

In order that a particular system described by Eq.~(\ref{phi2}) experience
a phase transition driven by a spontaneous symmetry breaking of the order
parameter from $m=0$ to $m\neq 0$, the symmetry that leads to $m=0$
must be embedded in the model. Note that if
Eq.~(\ref{phi2}) is invariant under the combined transformation 
\begin{eqnarray}
\varphi &\leftrightarrow &-\varphi ,
\notag\\
\xi &\leftrightarrow &-\xi ,  
\label{transf}
\end{eqnarray}
then $m=0$, the symmetric state, is indeed always a solution for
the order parameter. Equation~\eqref{phi2} satisfies the required symmetry
if $V(\varphi) $ and $\Gamma(\varphi) $ are \emph{even} functions.
We will thus consider this case
throughout the paper. Furthermore, we also assume with no loss of
generality that 
\begin{eqnarray}
V(0)&=&0, \notag \\
\Gamma(0)&=&1.  
\label{v0g0}
\end{eqnarray}

The exact stationary probability density of Eq.~(\ref{phi2}) can be
calculated for any noise interpretation, including the It$\hat{\rm o}$
and the Stratonovich interpretations~\cite{carrillo}.  However, any
further analytic insights require further approximation.  We
implement a \emph{mean-field approximation} in
Eq.~(\ref{phi2}) by replacing the average
value of the fields of the $2d$ nearest-neighbors of any site $i$
by the mean field value $\left\langle \varphi \right\rangle $, that is, 
\begin{equation}
\frac{1}{2d}\sum_{\langle ij \rangle}\varphi_j\rightarrow
\left\langle \varphi \right\rangle .
\end{equation}
This procedure, which is equivalent to assuming global coupling rather
than nearest neighbor coupling,
disregards fluctuations of the neighboring sites around the mean value.
Since all sites are then equivalent,
the lattice index can be dropped and the set of field equations reduces to
a single equation. However, the unknown mean value of the field appears in
this equation and must be chosen
\emph{self-consistently}. Thus, we obtain a closed approximate version of
the problem as expressed in the two equations
\begin{eqnarray}
\dot{\varphi}(t) &=&\Gamma (\varphi) \left( -\frac{
\partial V(\varphi)}{\partial \varphi }+K(\langle\varphi\rangle
-\varphi) \right) +\left[\Gamma(\varphi)\right]^{\frac{1}{2}}\xi(t),
\label{mf} \\ \nonumber\\
\langle \varphi \rangle &=&\langle \varphi \rangle_\rho . 
\label{mf2}
\end{eqnarray}
Here $\left\langle \cdot \right\rangle _{\rho }$ stands for a statistical
average with respect to the stationary probability density associated with
Eq.~\eqref{mf}, 
\begin{equation}
\rho_{\mathrm{st}}(\varphi;\langle\varphi\rangle)
=N(\langle\varphi\rangle) \Gamma(\varphi)^{(\alpha -1)}
e^{-\frac{2}{\sigma^2}\left( V(\varphi) +\frac{K}{2}(\langle\varphi\rangle
-\varphi)^2\right)},
\end{equation}
$N(\langle\varphi\rangle)$ is the normalization constant, and $\alpha=0$
$(\alpha= 1/2)$ for the It$\hat{\rm o}$ (Stratonovich) interpretation of
the noise.  This mean field formulation can not be solved in full
generality either, but it does allow some analytic characterization
of the problem.  It is this characterization that we pursue as far as
possible.

Note that the disordered solution (symmetric state) $\langle \varphi
\rangle =0$ always solves Eq.~(\ref{mf2}). Yet, other solutions such
that $\langle \varphi \rangle \neq 0$ are also possible. We refer
to the latter as ordered solutions. Note that as a consequence of the parity
of the functions $V(\varphi) $ and $\Gamma(\varphi)$,
$\langle\varphi\rangle_\rho$ is an odd function
of $\langle\varphi\rangle$ and therefore, if $\langle\varphi\rangle$
is a solution of Eq.~(\ref{mf2}), then so is $-\langle\varphi\rangle$.
However, both lead to the same value of the order parameter $m$.

At this point, we must make a distinction between \emph{solutions}
and \emph{phases} as determined by the stability of the former.
We call a \emph{disordered} (\textit{D}) phase a
macroscopic state where $\langle\varphi\rangle=0$ is the
\emph{only} stable solution.  If the solution
$\langle\varphi\rangle=0$ is unstable, and only a solution
with $\langle\varphi\rangle\neq 0$ is stable, the phase will be
called \emph{ordered} (\textit{O}).
If $\langle\varphi\rangle =0$
coexists with other stable, but ordered, solutions, the phase will be
denoted as \emph{multistable} (\textit{M}).

Since the solution of the self-consistency equation~(\ref{mf2}) determines
the acceptable values of $\langle\varphi\rangle$, it
is crucial to understand the behavior of $\langle\varphi\rangle_\rho$
as a function of $\langle\varphi\rangle $. Noting that
\begin{equation}
\frac{\partial \langle \varphi \rangle_\rho }{\partial
\langle \varphi \rangle }=\frac{2K}{\sigma^2}\left(
\langle \varphi^2\rangle_\rho-\langle \varphi
\rangle_\rho^2\right),
\label{slope}
\end{equation}
and applying the generalized Schwarz inequality 
\begin{equation}
\langle f^2(\varphi)\rangle_\rho \langle g^2(\varphi)
\rangle_\rho\geqslant \left|\langle f(\varphi)g(\varphi)
\rangle_\rho\right|^2  \label{sch}
\end{equation}
with $f(\varphi)=\varphi$ and $g(\varphi)=1$, 
one concludes that the right hand side of Eq.~\eqref{slope} is positive,
and therefore $\langle\varphi\rangle_\rho$ is a
monotonically increasing function of $\langle \varphi \rangle$. 
Moreover, taking the limit of 
\begin{equation}
\langle \varphi^n\rangle_\rho = \int_{-\infty}^\infty d\varphi\;
\varphi^n \rho_{\mathrm{st}}(\varphi;\langle\varphi\rangle)
\end{equation} 
as $\langle\varphi\rangle\to\pm\infty$ immediately leads to 
\begin{equation}
\lim_{\langle\varphi\rangle\rightarrow \pm\infty}\langle
\varphi^n\rangle_\rho
\to \langle\varphi\rangle_\rho^n,
\end{equation} 
and, consequently,
\begin{equation}
\lim_{\langle\varphi\rangle \rightarrow \pm \infty }\frac{
\partial\langle\varphi\rangle_\rho}{\partial\langle\varphi\rangle}=0.
\end{equation}
Therefore it follows that 
\begin{equation}
\lim_{\langle\varphi\rangle \rightarrow \infty}
\langle\varphi\rangle_\rho < \langle\varphi\rangle ,  \qquad
\lim_{\langle\varphi\rangle \rightarrow -\infty}
\langle\varphi\rangle_\rho > \langle\varphi\rangle , 
\label{limit}
\end{equation}
that is, $\langle\varphi\rangle_\rho$ necessarily lies below (above)
$\langle\varphi\rangle$ as $\langle \varphi\rangle$ goes to plus (minus)
infinity. 

\begin{figure}
\includegraphics[width = 6cm]{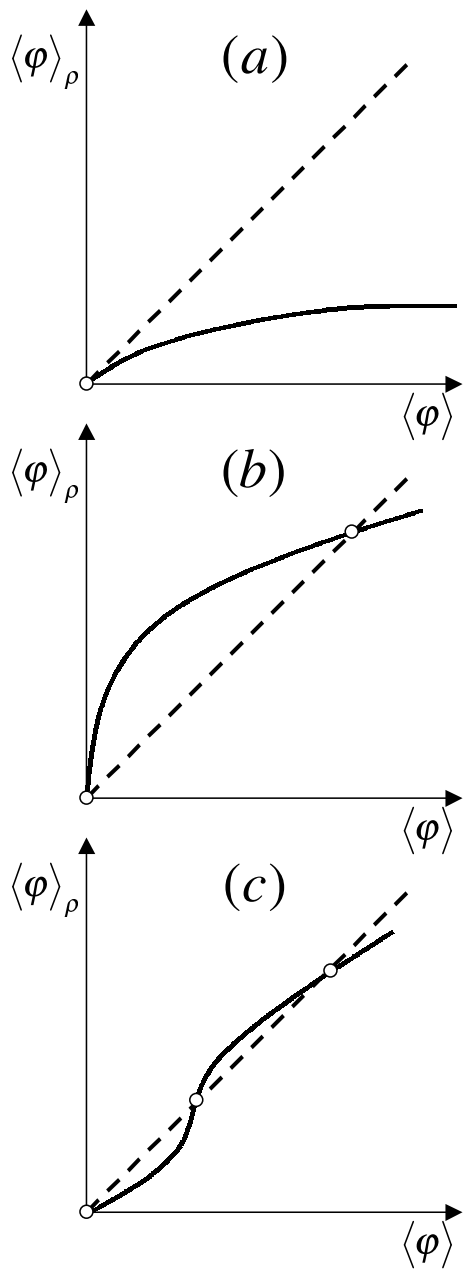}
\caption{Schematic of
the possible solutions for the self-consistency equation~(\ref{mf2}). If the
only solution is $\left\langle \protect\varphi \right\rangle =0$, panel (a),
the
system is in a \emph{disordered} phase. In panel (b), there is another
solution in addition to the one at
$\langle\varphi\rangle=0$. Only the nonzero
solution is stable, so that this represents
an \emph{ordered} phase. In panel (c) there is an
even number of nonzero solutions (here, two)
in addition to the $\langle\varphi\rangle=0$
solution.  The $\langle\varphi\rangle=0$ solution is stable, the next
intersection is unstable, and the third is again a stable solution.
The system thus exhibits \emph{multistability} in this
case, and involves the coexistence of a disordered and an ordered
solution.}
\label{f1}
\end{figure}

Figure~\ref{f1} illustrates the resulting possible
different phases in terms of the possible solutions of the self-consistency
equation~(\ref{mf2}). The dashed lines
represent $\langle\varphi\rangle$, the solid curves
$\langle\varphi\rangle_\rho$, and the self-consistency solutions are their
points of intersection.  Our subsequent discussions presume that
the system \emph{definitely} goes to a stable state.  We exclude
``runaway'' systems that do not fall into this category. 

\section{Phase Boundaries: Second and First Order Phase Transitions}
\label{boundaries}

With Fig.~\ref{f1} in mind, consider now the possible resulting
behaviors of the order parameter as we transition from one phase to
another by changing a control parameter. 
Note that
the general model~\eqref{phi2} (as well as the mean field version of the
model) depends on only two parameters, the coupling coefficient $K$ and
the noise intensity $\sigma^2$, so the control parameter that
characterizes a change from one phase to another could be either
of these two (or some combination of them).  The question then is how
the points of intersection in Fig.~\ref{f1} move as one varies a control
parameter that takes the system from the behavior shown in one panel to
that shown in another.

\begin{figure}
\includegraphics[width = 6cm]{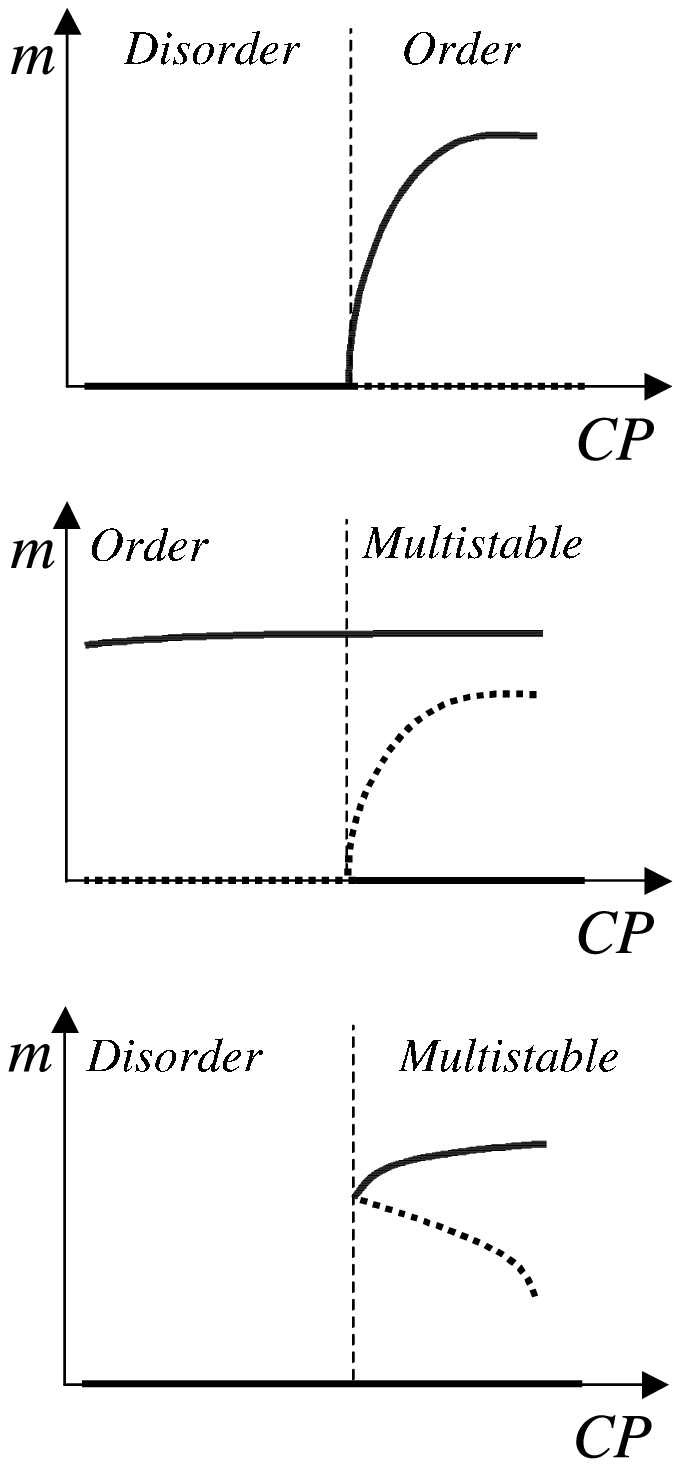}
\caption{Behavior of the order parameter, $m$, as a function of a
control parameter, $CP$, in the vicinity of a transition point. Transitions
between disordered and ordered phases are always second order (top
panel).  As shown in the middle and bottom panels, phase transitions
from, or to, multistable phases are (with the exception of singular
points discussed in the Appendix) first order.}
\label{f2}
\end{figure}

Consider the transition as a system moves from the behavior
shown in panel (a) of Fig.~\ref{f1} to that of panel (b). The way 
this is expected to occur is that the solid curve rotates upward so that
in addition to the $\langle\varphi\rangle=0$ solution, another solution
emerges at the origin. This second solution then moves upward along
the diagonal as the
control parameter increases.  This (a)$\to$(b) transition is illustrated
in the top panel of Fig.~\ref{f2}, where the solid lines represent
stable solutions and the dotted lines the unstable solutions.  The
transition between the disordered and ordered phases is continuous
in the order parameter (second order phase transition).  
A transition from panel (b) in Fig.~\ref{f1} to panel (c) would involve
the evolution of a kink in the curve that first cuts the diagonal
at the origin and then moves upward toward
the existing cut.  Associated with this there
is a change in the curvature of
$\langle\varphi\rangle_\rho$ near the origin.
This transition is sketched in the middle panel of Fig.~\ref{f2}, where
we show the two nonzero solutions moving closer together as the control
parameter increases.  The
transition between the ordered and multistable phases is discontinuous
(first order phase transition) and is expected to exhibit hysteresis.
The discontinuity is clearly apparent, for example, in the jump from the
disordered branch of the multistable phase as one decreases the control
parameter and this branch becomes unstable.
Finally, a transition from panel (a) to panel (c), the bottom panel of
Fig.~\ref{f2}, again involves a
change in the curvature near the origin and the evolution of a kink that 
first touches the diagonal at a single nonzero value that then separates
into two as the control parameter increases.  The transition between
disordered and multistable phases is again discontinuous.  Note that all
of these transitions can also proceed in the opposite direction than
illustrated here, e.g. a transition from order to disorder would occur
as in the top panel but from right to left.

We stress that the transitions between disordered and ordered phases are
continuous in the order parameter (second order phase transitions),
whereas the transitions from or to multistable phases are discontinuous
(first order) and are therefore expected to exhibit hysteresis. There
may be singular isolated exceptions to this latter conclusion, as
discussed in the Appendix.

A more quantitative characterization of the phase transitions is possible
for those transitions that involve a change in the stability properties of
the $\langle\varphi\rangle_\rho=0$ solution, that is, for phase transitions 
between disordered and ordered phases, and between ordered and multistable
phases (i.e., the top and middle panels in Fig.~\ref{f2}).
This characterization involves the first two nonzero derivatives
of $\langle\varphi\rangle_\rho$ with respect to $\langle\varphi\rangle$
in the vicinity of the origin.
The first derivative provides information about
the slope of $\langle\varphi\rangle_\rho$ and the third about the
concavity/convexity ($\langle\varphi\rangle_\rho$ is an odd function of
$\langle\varphi\rangle$, so even derivatives around the origin vanish).  
Transitions between disordered and ordered phases
(second order) occur when
\begin{eqnarray}
\left. \frac{\partial\langle\varphi\rangle_\rho}{\partial
\langle\varphi\rangle}\right|_{\langle \varphi
\rangle =0} &=&1,  \notag\\ 
\left. \frac{\partial^3\langle\varphi\rangle_\rho}{
\partial\langle\varphi\rangle^3}\right|_{\langle
\varphi\rangle=0} &<&0.  
\label{do}
\end{eqnarray}
Transitions between ordered and multistable phases (first order)
occur when
\begin{eqnarray}
\left. \frac{\partial\langle\varphi\rangle_\rho}{\partial
\langle\varphi\rangle}\right|_{\langle \varphi
\rangle =0} &=&1, \notag\\  
\left. \frac{\partial^3\langle\varphi\rangle_\rho}{
\partial\langle\varphi\rangle^3}\right|_{\langle
\varphi\rangle =0} &>&0.  
\label{om}
\end{eqnarray}
In turn, these derivatives can be expressed in terms of the cumulants of
the probability density $\rho_{\rm st}(\varphi;0)$. We proceed to establish
this relation.

Since $\langle\varphi\rangle_\rho$ is an odd function of
$\langle\varphi\rangle$, its
Taylor expansion around the solution
$\langle\varphi\rangle =0$ reads
\begin{equation}
\langle\varphi\rangle_\rho =\sum_{n=0}^{\infty }a_{2n+1}\langle
\varphi\rangle^{2n+1}, 
\label{g} 
\end{equation}
where
\begin{equation}
a_{2n+1} =\frac{1}{(2n+1)!}\int_{-\infty}^\infty  \left. \frac{
\partial^{2n+1}\rho_{\mathrm{st}}(\varphi;\langle \varphi
\rangle)}{\partial\langle\varphi\rangle^{2n+1}}
\right|_{\langle\varphi\rangle =0}\, \varphi \, d\varphi .  
\end{equation}
Furthermore, a straightforward but tedious calculation leads to the
result 
\begin{equation}
a_{2n+1}=\frac{2^{2n+1}}{(2n+1)!}\left(\frac{K}{\sigma ^{2}}
\right)^{2n+1}C_{2n+2},
\end{equation}
where $C_{2n+2}$ is the $(2n+2)^{\rm th}$ cumulant of the probability
distribution $\rho_{\mathrm{st}}(\varphi;0)$. The relation of
the cumulants to the statistical moments of the probability distribution
is given by 
\begin{equation}
C_{2n+2}=-\left| 
\begin{array}{ccccccc}
0 & 1 & 0 & 0 & 0 & 0 & \ldots \\ 
\langle\varphi^2\rangle_0 & 0 & 1 & 0 & 0 & 0 & \ddots \\ 
0 & \langle\varphi^2\rangle_0 & 0 & 1 & 0 & 0 & \ddots \\ 
\langle\varphi^4\rangle_0 & 0 & \binom{3}{1}\langle
\varphi^2\rangle_0 & 0 & 1 & 0 & \ddots \\ 
0 & \langle\varphi^4\rangle_0 & 0 & \binom{4}{2}
\langle\varphi^2\rangle_0 & 0 & 1 & \ddots \\ 
\langle\varphi^6\rangle_0 & 0 & \binom{5}{1}\langle
\varphi^4\rangle_0 & 0 & \binom{5}{3}\langle \varphi
^2\rangle_0 & 0 & \ddots \\ 
\vdots & \ddots & \ddots & \ddots & \ddots & \ddots & \ddots
\end{array}
\right|_{2n+2},
\end{equation}
where $\left|\cdot\right|_{2n+2}$ indicates the determinant of the
$(2n+2)\times(2n+2)$ matrix, $\binom{\cdot }{\cdot 
}$ the binomial coefficients, and $\langle \cdot \rangle_0$
stands for statistical averages over the probability density
$\rho _{\mathrm{st}}(\varphi;0)$.
Therefore, around $\langle \varphi \rangle =0$ the
self-consistency equation reads, 
\begin{equation}
\langle \varphi \rangle = \langle \varphi
\rangle_\rho
=\langle \varphi \rangle
\sum_{n=0}^{\infty }\frac{2^{2n+1}}{(2n+1)!}\left( \frac{K}{
\sigma^2}\right) ^{2n+1}C_{2n+2}\langle \varphi \rangle ^{2n} ,
\end{equation}
and it then follows that
\begin{eqnarray}
\left. \frac{\partial \langle \varphi \rangle_\rho}{\partial
\langle \varphi \rangle }\right|_{\langle \varphi
\rangle =0} &=&\frac{2K}{\sigma^2}C_2,  \\
\left. \frac{\partial^3\langle \varphi \rangle_\rho}{
\partial \langle \varphi \rangle^3}\right| _{\langle
\varphi \rangle =0} &=&8\left(\frac{K}{\sigma^2}\right)^3C_4.  
\end{eqnarray}

In summary, the boundaries between disordered and ordered phases and
between ordered and multistable phases are characterized by the first
two nonzero cumulants of the probability distribution
$\rho_{\mathrm{st}}(\varphi;0)$ as follows:
\begin{eqnarray}
\left. 
\begin{tabular}{c}
$C_2=\frac{\sigma^2}{2K}$ \\ 
$C_4<0$
\end{tabular}
\right\} &\Longrightarrow &\text{Second order phase transition
(order-disorder boundary)} \label{condspt}\\
\left. 
\begin{tabular}{c}
$C_2=\frac{\sigma^2}{2K}$ \\ 
$C_4>0$
\end{tabular}
\right\} &\Longrightarrow &\text{First order phase transition
(order-multistable boundary).}
\label{condspt2}
\end{eqnarray}
Transitions between disordered and multistable phases can not be
characterized in this fashion since they require information about
$\langle \varphi \rangle_\rho$ away from $\langle \varphi \rangle=0$.

\section{Overall Structure of Phase Diagrams}
\label{overall}

Having discussed the possible phase transitions that might be observed
in the mean field system \eqref{mf}-\eqref{mf2}, we now ask which
particular phases might be present for particular values of the control
parameters $K$ and $\sigma^2$.  In this section we present an analytic
deduction of the phases present for small and for large values of the
coupling coefficient $K$.  The behavior for intermediate values must be
deduced on the basis of plausibility arguments that we introduce later.

We start by defining
\begin{equation}
\Theta_K=a_1-1=\frac{2K}{\sigma^2}C_2-1.
\end{equation}
This quantity measures the differences
in the slopes of $\langle \varphi \rangle $ and $\langle
\varphi \rangle_\rho$ as a function of $\langle \varphi \rangle $
near the origin.  According to the analysis presented in the previous
section, $\Theta_K>0$ for an ordered phase and $\Theta_K<0$ in either
disordered or multistable regions of the phase diagram. Moreover, the phase
boundaries to or from ordered states are given by the zeros of $\Theta_K$.

Note that $\Theta_{0}=-1$, that is, $\langle \varphi \rangle_\rho=0$
in the absence of coupling. Thus, $\langle \varphi
\rangle =0$ is the only possible solution to the self-consistency
equation~(\ref{mf2}) in the small coupling limit, and the system is
disordered in this limit. \emph{At sufficiently weak coupling the system
is therefore always disordered.}

One can easily check that 
\begin{equation}
\left. \frac{\partial \Theta_K}{\partial K}\right|_{K=0}>0.
\end{equation}
Thus, as $K$ grows from zero the system advances toward
the ordered phase. This statement does not mean that the system will
actually enter into the ordered phase as the coupling increases;
it simply states the ordering role of weak but increasing coupling.

On the other hand and more interestingly, it is possible to investigate
the strong coupling limit as follows. We first introduce the 
convenient notation 
\begin{eqnarray}
\Phi (\varphi) &=&\Gamma(\varphi)^{\alpha -1}e^{-
\frac{2}{\sigma^2}V(\varphi)}, \\ \nonumber \\
I_{2n}(K,\sigma^2) &=&\int_{-\infty}^\infty \varphi^{2n}\;\Phi
(\varphi)e^{-\frac{K}{\sigma^2}\varphi^2}\,d\varphi ,
\end{eqnarray}
where $n\geqslant 0$. We can then write all the 
non-zero moments of $\rho_{\mathrm{st}}
(\varphi;0)$ as $\langle \varphi^{2n}\rangle_0=I_{2n}/I_0$.
In particular, the cumulants of interest here can be written as
\begin{equation}
C_2=\frac{I_2}{I_0}, \qquad C_4= \frac {I_4}{I_0} -3\frac{I_2^2}{I_0^2}.
\label{cums}
\end{equation}
Moreover, notice that all the
moments can be reduced to the calculation of $I_0$ since 
\begin{equation}
I_{2n}(K,\sigma^2)=(-1)^{n}\sigma^{2n}\frac{\partial^n I_0(K,\sigma^2)}
{\partial K^n}.
\label{i2n}
\end{equation}
A series expansion of $I_0$ useful for large values of $K$
follows from an expansion of
$\Phi(\varphi)$ around $\varphi=0$, which allows us to carry out the
integral:
\begin{equation}
I_{0}(K,\sigma^2) =\int_{-\infty}^\infty
\sum_{m=0}^{\infty }\left( \frac{
\Phi ^{(m)}}{m!}\varphi^{m}\right) e^{-\frac{K}{\sigma ^{2}}\varphi
^{2}}d\varphi =\sum_{m=0}^{\infty }\frac{\Phi ^{(2m)}\pi^{1/2}}{
m!\,2^{2m}}\left( \frac{\sigma ^{2}}{K}\right)^{m+1/2}, 
\label{io}
\end{equation}
where, 
\begin{equation}
\Phi ^{(z)}=\left. \frac{\partial^z\Phi(\varphi)}{\partial\phi^z}
\right|_{\varphi =0}.
\end{equation}
Introducing Eq.~(\ref{io}) into Eq.~(\ref{i2n}), we obtain the series
\begin{equation}
I_{2n}(K,\sigma^2)=
\sum_{m=0}^{\infty }\frac{\Phi ^{(2m)}\pi^{1/2}\left[ m+1/2\right]_{n}}
{m!\,2^{2m}}\left( \frac{\sigma
^{2}}{K}\right)^{m+n+1/2},
\end{equation}
where 
\begin{equation}
\left[ z\right]_{n} =\prod_{l=0}^{n-1}(z+l), \qquad
\left[ z\right]_{0} =1 .
\end{equation}
The moments of $\rho_{\mathrm{st}} (\varphi;0)$ then read 
\begin{equation}
\langle\varphi^{2n}\rangle_0 =
\left( \frac{
\sigma^2}{K}\right)^n\frac{\displaystyle \sum_{m=0}^{\infty }
\frac{\Phi ^{(2m)}}{m!\,2^{2m}}\left[ m+1/2\right]_{n}\left(\frac{\sigma^2}
{K}\right)^m}{\displaystyle \sum_{m=0}^{\infty }\frac{\Phi^{(2m)}}{m!\,2^{2m}}
\left(\frac{\sigma^2}{K}\right)^m}. 
\label{phi2n}
\end{equation}

A more convenient expression for Eq.~(\ref{phi2n}) is obtained by performing
its Taylor expansion around $(\sigma^2/K)\rightarrow 0$, 
\begin{eqnarray}
\langle \varphi^{2n}\rangle_0&\underset{K\gg 1}{=}&\left( 
\frac{\sigma^2}{K}\right)^n\left[ \left[ 1/2\right] _{n}+\left( \frac{
\sigma^2}{K}\right)\frac{\Phi ^{(2)}}{8\Phi ^{(0)}}
\left( \left[ 3/2\right]_{n}-\left[1/2\right]_{n}\right)
\right.  \notag \\ \notag\\
&&+\left( \frac{\sigma^2}{K}\right)^2\frac{1}{32\left(\Phi^{(
0)}\right)^2}\left(\Phi^{(4)}\Phi^{(0)
}\left( \left[ 5/2\right]_{n}-\left[ 1/2\right]_{n}\right) \right.  \notag
\\ \notag \\
&&\left. \left. -2\left(\Phi^{(2)}\right)^2\left(\left[
3/2\right]_{n}-\left[1/2\right]_{n}\right) \right) +\mathcal{O}\left(
\left( \frac{\sigma ^2}{K}\right)^3\right) \right] .  
\label{momento2approx}
\end{eqnarray}
The first term in the series~(\ref{momento2approx}), i.e., up to
order $(\sigma^2/K)^n$, leads to the familiar result of applying
the \emph{steepest descent} method \cite{arfken} to $I_0$, 
\begin{equation}
\langle \varphi ^{2n}\rangle_0\underset{K\rightarrow \infty }{=
}\left(\frac{\sigma^2}{K}\right)^n\left[ 1/2\right] _{n}.
\label{sd}
\end{equation}
However, this result is not sufficiently accurate to capture enough of
the large-coupling behavior of $\langle\varphi^{2n}\rangle_0$
and shed light on the behavior of the phase boundaries in that limit.
Keeping up to the next order, that is,
\begin{equation}
\langle \varphi^{2n}\rangle_0\underset{K\gg 1}{=}\left(
\frac{\sigma^2}{K}\right)^n\left[ \left[ 1/2\right] _{n}+\left( \frac{
\sigma^2}{K}\right)\frac{\Phi ^{(2)}}{8\Phi ^{(0)}}
\left( \left[ 3/2\right]_{n}-\left[1/2\right]_{n}\right)
+\mathcal{O}\left( \left(
\frac{\sigma ^{2}}{K}\right)^{2}\right) \right],
\end{equation}
one finds for the function $\Theta_K$ 
\begin{equation}
\Theta _{K\gg 1}=\frac{1}{2}\left(\frac{\sigma^2}{K}\right)\frac{\Phi
^{(2)}}{\Phi^{(0)}}+\mathcal{O}\left( \left( 
\frac{\sigma ^{2}}{K}\right) ^{2}\right) .
\end{equation}
Note that $\Phi^{(0)}=1$ [c.f. Eq.~(\ref{v0g0})], and therefore
the sign of $\Theta_{K}$ for large values of the coupling is determined by
the sign of $\Phi^{(2)}$, 
\begin{equation}
\Phi ^{(2)}=(\alpha-1)\Gamma^{(2)}-
\frac{2}{\sigma^2}V^{(2)}. 
\label{G2}
\end{equation}
If $\Phi ^{(2)}>0$ then $\Theta_{K\gg 1}>0$ (ordered phase).
On the other hand, if $\Phi^{(2)}<0$ then $\Theta_{K\gg 1}<0$
(disordered or multistable phase). That is, whether or not the system is
in an ordered phase depends only on the balance of convexities
of the local potential and the field-dependent coefficient at the origin.

Furthermore, since $\Theta_{K\gg 1} = \mathcal{O}\left( \left( \sigma
^{2}/K\right) \right)$, which vanishes as $K$ increases, one knows that
for large coupling the system is ``near'' a phase boundary of an ordered
phase. One can gain some insight into the type of transition that might
be involved by studying the fourth cumulant, $C_4$ [cf.
Eqs.\eqref{condspt} and \eqref{condspt2}].
Using Eq.~(\ref{momento2approx}) in Eq.~\eqref{cums} and recalling 
that $\Phi^{(0) }=1$, we obtain, 
\begin{equation}
C_4\left( K\gg 1\right) =\frac{1}{16}\left( \frac{\sigma
^{2}}{K}\right)^4
\left( \Phi ^{(4)}-3\left( \Phi ^{(2)}\right)
^{2}\right) +\mathcal{O}\left( \left( \frac{\sigma ^{2}}{K}\right)
^{5}\right) ,
\end{equation}
where, 
\begin{eqnarray}
\Phi ^{(4)} &=&-\frac{12}{\sigma^2}(\alpha -1)
V^{(2)}\Gamma^{(2)}+\frac{12}{\sigma^4}\left(
V^{(2)}\right)^2+(\alpha -1) \Gamma ^{(4) } \notag \\
&&+3(\alpha -1)(\alpha -2) \left( \Gamma^{(2)}\right)^2
-\frac{2}{\sigma^2}V^{(4)} .
\end{eqnarray}
We thus confirm that, independently of the behavior of the system at
intermediate values of the coupling, for large coupling the
appearance or disappearance of ordered
phases as reflected in the sign of the fourth cumulant
depends on the geometrical properties of $V(\varphi)$ and of
$\Gamma(\varphi) $ around the origin.

\begin{figure}
\includegraphics[width = 9cm]{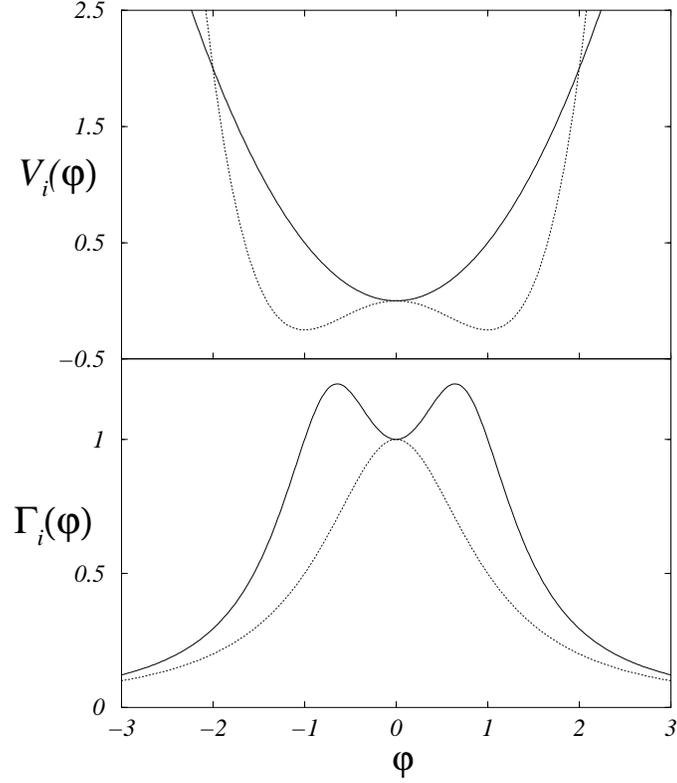}
\caption{Generic local
potentials $V_i(\varphi)$ and field-dependent coefficients
$\Gamma_i(\varphi)$ as a function of the
field $\varphi$. The solid lines are for $i=1$ and 
the dotted lines for $i=2$. The behavior at the origin of the
derivatives of $V(\varphi)$ and $\Gamma(\varphi)$ determine the
phase at large values of the coupling (see text).}
\label{f4}
\end{figure}

There are basically only two distinct generic types of behavior of these
functions around the origin, and therefore only four possible
combinations.  The possible types of functions are shown in
Fig.~\ref{f4}, where we have plotted the simple representative
cases~\cite{nota1} 
\begin{eqnarray}
V_1(\varphi)&=&\frac{\varphi^2}{2},
\notag \\
V_2(\varphi) &=&\frac{\varphi^4}{4}-\frac{\varphi^2}{2}
\label{vs}
\end{eqnarray}
and
\begin{eqnarray}
\Gamma_1(\varphi)&=&\frac{ 1+\varphi^2} {1+\varphi^4}, \notag\\
\Gamma_2(\varphi)&=&\frac{1}{1+\varphi^2} .
\label{gs} 
\end{eqnarray}
Note that the field-dependent coefficient $\Gamma_1$ favors fluctuations
around $\varphi=\pm1$ while $\Gamma_2$ leads to the largest fluctuations
around $\varphi=0$~\cite{marta}.
It is quite straightforward to determine the sign of $C_4$ on the basis
of the derivatives of these functions (one can use the generic forms
\eqref{vs} and \eqref{gs} as a guide) and compile the following table.
We emphasize that \emph{these are results in the strong coupling limit}.
The entries in the table indicate order (\textit{O}), disorder
(\textit{D}), and multistability (\textit{M}):
\begin{equation*}
\begin{tabular}{|c|c|c|}
\hline
& $V_1(\varphi)$ & $V_2(\varphi)$ \\ \hline
$\Gamma_1(\varphi)$ & \textit{M} & $
\begin{array}{c}
\text{\textit{O} if }\sigma^2<\sigma_c^2 \\ 
\text{\textit{M} if }\sigma^2>\sigma_c^2
\end{array}
$ \\ \hline
$\Gamma_2(\varphi) $ & $
\begin{array}{c}
\text{\textit{O} if }\sigma^2>\sigma_c^2 \\ 
\text{\textit{D} if }\sigma^2<\sigma_c^2
\end{array}
$ & \textit{O} \\ \hline
\end{tabular}
\end{equation*}
Here $\sigma_c^2$ is a critical value of the noise intensity that
separates different phases; for the generic models displayed above,
$\sigma_c^2=1/\left( 1-\alpha \right)$.
Note that the noise interpretation through the value of $\alpha $
simply shifts the critical value of $\sigma_c^2$. The
case $V_1$, $\Gamma_2$ with the functions given above
has recently been studied in the context of phase transitions and pattern
formation~\cite{marta, carrillo, buceta}. 
Also, in agreement with our general analysis it was noted
recently~\cite{carrillo}
that in that particular case the noise-induced phase transition is not
attributable to the so-called Stratonovich drift, as is the case in other
noise-induced phenomena~\cite{chris}.

We have thus arrived at the generic phase structure for the mean field
problem \eqref{mf}-\eqref{mf2} in the weak coupling limit (the system is
disordered) and in the strong coupling limit (as shown in the table).
For intermediate coupling we are not able to provide a general
quantitative analysis except to note that if $\sigma ^{2}\rightarrow 0$,
the system is always in a disordered (\textit{D}) phase 
since the fluctuations are needed to provide the energy to induce
symmetry breakings.

Despite this difficulty, one can introduce compelling arguments to
connect the phase behavior that we have established in these limits, and
to arrive at a set of full phase diagrams.

There are several unknown regions connecting various known
phases at this point.  In the $V_1,\Gamma_1$ case we need to connect
the weak-coupling \textit{D} phase to the strong-coupling
\textit{M} phase. For the $V_2,\Gamma_1$
combination we require a connection between the \textit{D} phase and an
\textit{O} phase if the noise is weak,
or to an \textit{M} phase if it is
strong.  With the $V_1,\Gamma_2$ combination the connection needs to be
made between the disordered weak-coupling phase and an ordered or a
disordered strong-coupling phase depending on the noise intensity.  And
in the $V_2,\Gamma_2$ case a connection needs to be established from the
disordered to the ordered phase.  The simplest
possible scenarios for connections are the following.  The simplest
connection between disordered phases is simply a disordered phase, i.e.,
a situation where no phase transition occurs at all.  A connection
between disordered and ordered phases is most straightforwardly
accomplished through a single second-order phase transition.
Finally, for the connection between disordered and
multistable phases, two different scenarios are most feasible. 
One possibility is that the 
connection is mediated through an ordered phase, as follows. 
As noted in Fig.~\ref{f2}, when a multistable
region appears from a disordered phase, the unstable solution tends
at first to move downward as the control parameter increases. If
the unstable solution eventually vanishes,
the disordered phase necessarily becomes unstable and one
necessarily enters an ordered phase. 
Such destabilization does not occur if the transition is mediated by and
ordered phase.  On the contrary, multistable phases arising from
\textit{OM} transitions grow more stable as the control parameter
increases (see Fig.~\ref{f2}).
In this case, a feasible sequence would be of
the form \textit{DOM}.  On the other hand,  a direct 
\textit{DM} transition may also occur, but only if the initial
vanishing tendency of the unstable solution is stabilized as the
coupling increases.

\begin{figure}
\includegraphics[width = 10cm]{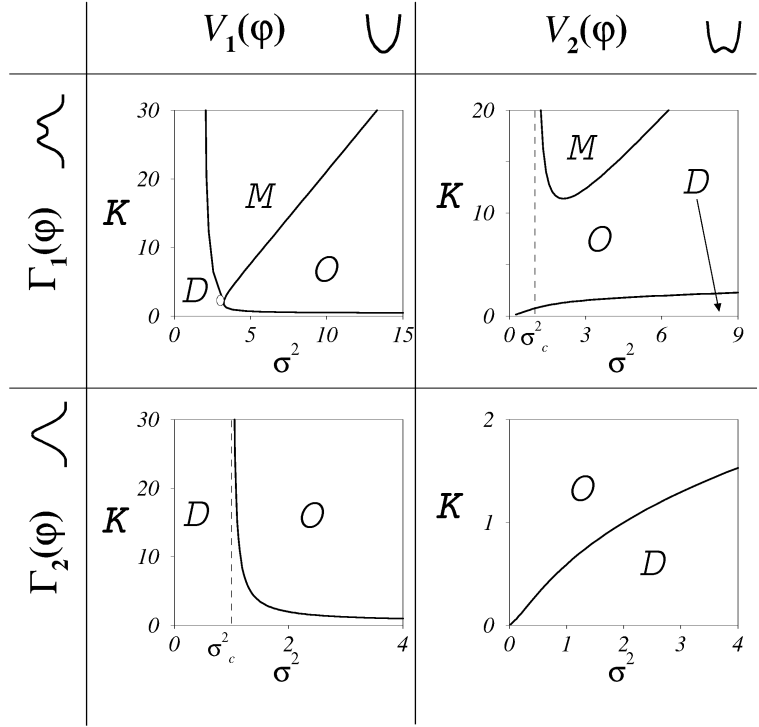}
\caption{Mean-field phase diagrams as a
function of the local potentials and field-dependent kinetic coefficients
given by Eqs.~(\ref{vs}) and~(\ref{gs}).
The small
open circle in the phase diagram for $\Gamma_1$ and $V_1$ where the
three phases merge indicates an isolated singular critical point
(\emph{triple} point) where a continuous phase transition between
disordered and ordered phases occurs (see the appendix). The
overall structure of the phase diagrams is in perfect agreement with
the schematic structures discussed in the text.}
\label{f6}
\end{figure}

We can corroborate this scenario by calculating the phase diagrams that are
obtained from the mean field approximation for the particular
functions \eqref{vs} and \eqref{gs}. We numerically solve the
self-consistency equation~(\ref{mf2}) and compute the boundaries separating
different phases. Recall that the noise interpretation simply shifts the
transitions points but does not change the phase diagram structure. We
present the results for $\alpha =0$, that is, the It$\hat{\rm o}$
interpretation, for which $\sigma_c^2=1$.  The results are shown
in Fig.~\ref{f6}.

Note that the actual structure of the phase diagrams at small and large
values of the couplings is perfectly captured by our analysis. Moreover, the
value of the critical noise intensity is $\sigma_{c}^{2}=1$, as predicted.
As for the unknown domains shown for intermediate coupling, our
arguments about the simplest scenarios agree with the mean-field
results. For example, for the $V_2,\Gamma_1$ combination
the appearance of the multistable phase with increasing $K$
is mediated through an ordered
phase. We point out that for the case $V_1,\Gamma_1$, there is
a \emph{triple} point where all phases merge. At this critical point
a continuous phase transition between disordered and multistable
phases occurs. As noted earlier, this behavior is singular
and isolated.  Note also that multistability
appears in this case by means of a \textit{DOM} sequence
with increasing coupling $K$ for noise intensities to the
right of the triple point. However, to the left of the triple point the
multistable phase arises from a \textit{DM} sequence.  Moreover, above
the triple point and with increasing $\sigma^2$, multistable phases are
always destabilized and followed by an ordered phase, as mentioned
above. There is
evidently an asymptote at $\sigma^2\simeq 2$
for the phase boundary separating the disordered and
multistable phases.  However, this critical
noise intensity is not captured by our theory since it
does not involve slopes and convexities \emph{near} the origin.
Instead, we show below how this second critical value of the noise
intensity can be calculated by analyzing the zero-dimensional version of the
problem. 

A number of other striking features of the phase diagrams are
noteworthy.  For both $V_1$ cases, independently of the value of
the coupling, \emph{the system becomes more ordered as the noise intensity
increases}. In the case of $\Gamma_1$ this behavior is associated with
the destabilization of multistable phases and in general
suggests that the phenomenon of
the so-called \emph{inverted phase diagrams} depends mainly on
the convexity of the local
potential around the origin. It is also worth noting the
phenomenon of reentrant noise-induced multistability for the
case $V_2,\Gamma_1$. If the coupling is greater than
$K\sim 10$, increasing the noise intensity causes a transition
from an ordered phase to a multistable phase. However a further
increase in the noise intensity eventually leads the system back to 
the ordered phase. This behavior resembles the phenomenon of
reentrance as a function of the intensity of the fluctuations
in other noise-induced phenomena~\cite{chris, buceta2}.
However, in the latter the phase changes are from disordered to ordered
and, for sufficiently intense noise, back to disorder.

\emph{Zero-Dimensional Analysis.}
Previous studies for the particular case $V_1$, $\Gamma_2$~\cite{marta,
carrillo} have revealed that in the case of relaxational-flows with field
dependent relaxation coefficients the mechanism responsible for the phase
transition is similar to that which drives the noise-induced
transition \emph{a la} Horsthemke-Lefever in zero-dimensional
systems~\cite{horsthemke}. In the zero-dimensional case,
noise-induced transitions are associated with changes in the extrema of the
local potential. We present an analysis of the zero-dimensional system
to compare with some of the results of our mean-field analysis. The
zero-dimensional version of Eq.~(\ref{phi2}) reads
\begin{equation}
\dot{\varphi}(t) =\Gamma(\varphi)\left( -\frac{\partial
V(\varphi)}{\partial \varphi}\right) +\left[ \Gamma
(\varphi)\right]^{\frac{1}{2}}\xi(t) ,
\label{od}
\end{equation}
that is, the uncoupled version of our original model~\eqref{phi2}.
The stationary probability density now is
\begin{equation}
\rho _{\mathrm{st}}(\varphi)=Ne^{-\frac{2}{\sigma^2}V_{
\mathrm{eff}}(\varphi)} ,
\end{equation}
where $N$ is the normalization constant and
$V_{\mathrm{eff}}(\varphi) $ is the\emph{\ effective}
potential,
\begin{equation}
V_{\mathrm{eff}}(\varphi)=V(\varphi)+\frac{\sigma^2(1-\alpha)}{2}
\ln\Gamma(\varphi) .
\end{equation}
The equilibria of the effective potential are given by the
condition $V_{\mathrm{eff}}^{\prime}(\varphi^\ast)=0$,
that is,
\begin{equation}
\Gamma (\varphi^{\ast }) V^{\prime }(\varphi^{\ast})
+\frac{\sigma^2(1-\alpha)}{2}\Gamma^{\prime}(\varphi^{\ast })=0 . 
\label{equi}
\end{equation}
The stability of the equilibria depends on the sign of the second
derivative of the potential at the equilibrium points.
A noise-induced transition occurs when there is a
change in the stability of the solution $\varphi^{\ast }$. Therefore,
the boundary of stability is given by
$V_{\mathrm{eff}}^{\prime \prime }(\varphi^{\ast })=0$, that is,
\begin{equation}
\Gamma^2(\varphi^{\ast})V^{\prime \prime}(\varphi^{\ast })
+\frac{\sigma^2(1-\alpha)}{2}\left(\Gamma (\varphi^{\ast })
\Gamma ^{\prime \prime }(\varphi^{\ast })-\left(\Gamma^{\prime }
(\varphi ^{\ast })\right)^2\right)=0 . 
\label{esta}
\end{equation}

Note that $\varphi ^{\ast }=0$ is always a solution of Eq.~(\ref{equi}) and
therefore the disordered solution is stable if 
\begin{equation}
V^{\prime \prime }(0)+\frac{\sigma^2(1-\alpha)}{2}
\Gamma^{\prime \prime }(0)=0 .
\end{equation}
This equation corresponds exactly
to the stability boundary associated with 
Eq.~(\ref{G2}) . In other words, 
\emph{the strongly coupled system behaves exactly 
as the uncoupled system}. In the cases
$V_2(\varphi), \Gamma_1(\varphi)$ and $V_1(\varphi),
\Gamma_2(\varphi)$ the critical value of the noise intensity that
changes the stability of the disordered solution
to an ordered one is exactly as calculated in the coupled system, 
$\sigma_c^2=1/(1-\alpha)$.

Moreover, our purpose in analyzing the zero-dimensional system is
also to understand the phase boundary that
separates multistable and disordered phases in
the case $V_1,\Gamma_1$ in the spatially extended
problem.  Recall that in that case the
stability of the disordered solution does not change, and therefore
we are not able to use our mean field analysis near the disordered state to
compute phase boundaries.
However, we can use Eqs.~(\ref{equi}) and \eqref{esta} to
support our numerical findings that indicated that there is
a critical noise intensity
separating those two phases.  While solving Eq.~(\ref{equi})
for solutions $\varphi^{\ast }\neq 0$ is rather cumbersome and
Eq.~(\ref{esta}) does not have an analytic solution for those
values, it is trivial to solve the problem numerically. 
The result agrees perfectly with our previous findings: there is
an asymptote at $\sigma^2\simeq 2$ (It$\hat{\rm o}$) that
corresponds to the critical value of the noise intensity separating
disordered and multistable phases.

\section{Numerical Simulations}
\label{numerical}

To check the predictions of the mean field theory we present
numerical simulations of Eq.~(\ref{phi2}) in a two dimensional
square lattice with
nearest neighbor interactions, the It$\hat{\rm o}$
interpretation of the noise, and periodic
boundary conditions. We focus on the case $V_1,\Gamma_1$.
This is the most interesting, previously unexplored, case:
a striking effect of the noise, a noise-induced multistable
phase and an inverted phase diagram, occur in this case.
Note that the case $V_1,\Gamma_2$ has been studied recently~\cite{marta},
the case $V_2,\Gamma_2$ presents a phase diagram with a
phenomenology similar to the well-known model A~\cite{halperin,raul},
and the case $V_2,\Gamma_1$ presents as its main feature the same
striking phenomenology of noise-induced multistability as does the case
$V_1,\Gamma_1$.

\begin{figure}
\includegraphics[width = 9cm]{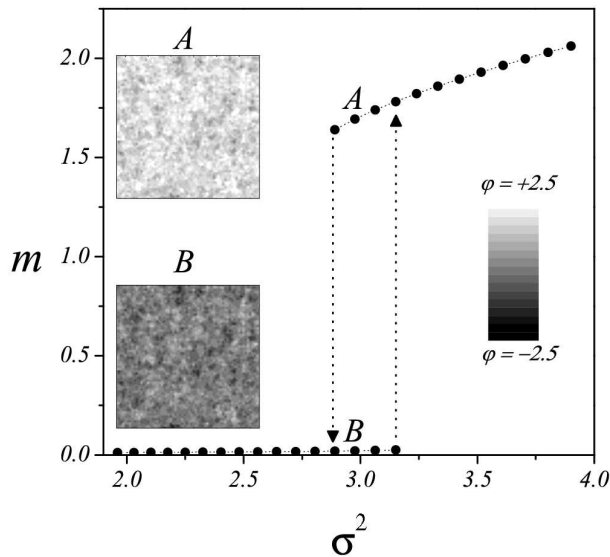}
\caption{Order parameter, $m$, as a function of the noise intensity, $\sigma^2$, for the
case $V_1,\Gamma_1$ and a fixed value of the coupling $K=10$.
The noise term is interpreted in the It$\hat{\rm o}$ sense.
Note that the system explores the three possible phases
(disordered, multistable, and ordered) as
the intensity of the noise is increased. The insets show the steady
states of the field for the points \textit{A} and \textit{B} indicated
in the figure (see text). The noise intensity at these points is
$\sigma^2\approx 3$.}
\label{f7}
\end{figure}

Figure \ref{f7} shows the order parameter $m$
as a function of the noise intensity $\sigma^2$ 
for a fixed value of the coupling constant, $K=10$.
The system is seen to explore the three possible phases,
disordered, multistable, and ordered, as
the fluctuations become stronger. Moreover, the system presents
an inverted phase
diagram where order becomes more prominent as the noise is increased.
As indicated by the discontinuous behavior of the order parameter, the
phase transitions are first order in all cases.
To detect the multistable phase and the associated hysteresis,
we integrated Eq.~(\ref{phi2}) under two different conditions.
The initial conditions
and the noise realizations are identical in both cases.
A difficulty in such simulations is that it takes an inordinately long time
to reach a steady state (eventually the system leaves any steady
state if the system is finite, but this time can be made as long as desired
by increasing the size of the system). 
To overcome this difficulty,
in one case we added a very small external field
that favors the solution $m=0$ while in the other we added one that
favors an ordered solution.
As soon as steady states were reached, the external fields were turned-off.
The insets of Fig.~\ref{f7} show, by means of a density plot, the
values of the
field in the multistable phase for the points \textit{A} and
\textit{B} for which $\sigma^2\approx 3$.
The scale of the density plots is also presented and is the
same for both insets.
The insets highlight the striking feature of the noise-induced
multistability and show the two possible states within the multistable region.

\section{Conclusions}
\label{conclusion}

We have presented a detailed studied of phase transitions in
models
with field-dependent relaxation coefficients. By means of
a mean-field approximation
in combination with other analytical techniques and plausibility 
arguments supported by particular examples, we have elucidated
the phase diagrams
that can be found in such models. We stress that our methods can easily
be applied to a variety of other systems.
Moreover, we have demonstrated that
disorder-multistability continuous phase transitions are singular points
in the phase diagram, and that the phases 
for large values of the coupling
are determined by geometrical features of the local potential and
of the field-dependent coefficient in the vicinity of
the origin. We have also showed that the
behavior of the extended system at large values of the coupling coefficient
is equivalent to the behavior of the uncoupled
zero-dimensional system. Therefore, the mechanism responsible for
the phase transitions
is similar to the noise-induced transitions \textit{a la}
Horsthemke-Lefever, and
is not attributable to the Stratonovich drift together with
collective effects involved in other
noise-induced phenomena~\cite{chris, buceta2}.
Finally, we have performed numerical simulations of a particular
case to check the results of the mean-field
approximation. The numerical results are in qualitative agreement
with the theoretical
predictions and reproduce the main features of the system, most notably,
the occurrence of noise-induced multistability,
and of an inverted phase diagram indicating that stronger noise induces
greater order.

We envision further modifications of these models that would
extend the richness of the observed phenomenology. In particular,
including other degrees of freedom and considering different
couplings increases the complexity of the multistability
phenomena caused by the noise. Further degrees of freedom
could, for example, lead to \textit{locking} of the system either in
an oscillatory mode (limit-cycle)
or in a stationary state (focus) depending on the initial conditions.
Considering couplings that favor morphological instabilities could lead to
pattern formation determined
entirely by the initial conditions. Work in these directions is
in progress.

\begin{acknowledgments}
This work was partially supported by the Engineering Research Program of
the Office of Basic Energy Sciences at the U. S. Department of Energy
under Grant No. DE-FG03-86ER13606,
by MECD-Spain Grant No. EX2001-02880680, and by MCYT-Spain Grant 
No. BFM2001-0291.
\end{acknowledgments}

\appendix
\section{Singular Transition}
\label{appendixa}

In Fig.~\ref{f2} and the discussion surrounding it we noted that with
the exception of singular points in the parameter space,
phase transitions from
disorder to multistability are first order (discontinuous).  In this
appendix we expand on this assertion.
By ``singular'' we mean that if there is a point $(\sigma_{\ast }^2,
K_{\ast }) $ in the phase diagram where a continuous transition between
disorder and multistability exists,
then no neighboring points in the phase diagram can
present the same transition. In other words, continuous
disorder-multistability phase transitions are \emph{isolated critical points}
in the phase diagram, and there is no plausible continuous function
$K(\sigma^2)$ connecting them.

\begin{figure}
\includegraphics[width = 8cm]{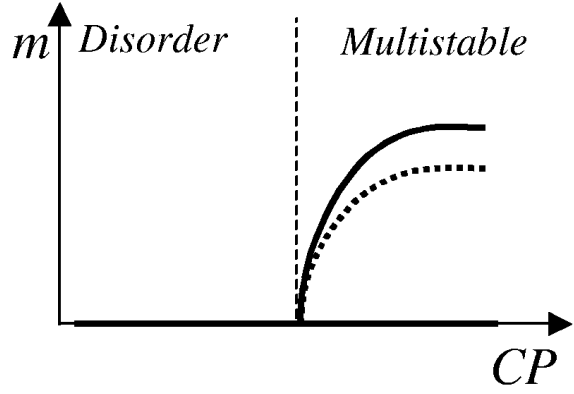}
\caption{Schematic behavior of the order
parameter as a function of a control parameter
in the vicinity of a continuous disorder-multistability phase
transition. This kind of behavior is singular and isolated in the phase
diagram.}
\label{f3}
\end{figure}

Figure~\ref{f3} shows the behavior of the order parameter $m$
as a function of a control parameter in the
vicinity of a continuous disorder-multistability
phase transition. Note that for
that behavior to happen, all three roots of the self-consistency
equation must vanish exactly at the critical value of the control
parameter. Moreover, the convexity
of $\langle \varphi \rangle_\rho$ must also change sign at 
exactly that value. If such a critical point $(\sigma
_{\ast }^2,K_{\ast })$ exists, it must satisfy the conditions
\begin{equation}
\left. 
\begin{array}{c}
C_2=\frac{1}{2}\left( \frac{\sigma_{\ast}^2}{K_{\ast }}\right) \\ \\
C_{4}=0
\end{array}
\right\} \Longrightarrow \left\{ 
\begin{array}{c}
\langle\varphi^2\rangle_0=\frac{1}{2}\left( \frac{\sigma
_{\ast }^2}{K_{\ast }}\right) \\ \\
\langle\varphi^4\rangle_0=\frac{3}{4}\left( \frac{\sigma
_{\ast }^2}{K_{\ast }}\right)^2
\end{array}
\right. .
\end{equation}
These requirements fulfill the Schwarz inequality, Eq.~(\ref{sch}), and we
must therefore conclude that such a critical point is possible.

Now, assume the existence of the critical point $(\sigma _{\ast}^2,
K_{\ast })$. We investigate the requirements for
a neighboring point $(\sigma_{\ast }^2+\varepsilon_{\sigma
^2},K_{\ast }+\varepsilon_K) $ to also be associated with a
disorder-multistability phase transition. A straightforward calculation
leads to
the following condition to be satisfied by $\varepsilon_{\sigma ^2}$, 
\begin{equation}
\varepsilon_{\sigma^2}\left(\langle\varphi^2\rangle_0
\langle V(\varphi)\rangle_0-\langle
\varphi^2V(\varphi)\rangle_0\right)=0 .
\end{equation}
This can in general only be satisfied if
$\varepsilon_{\sigma^2}=0$. As for $\varepsilon_K$,
it must satisfy
\begin{equation}
\varepsilon_K\left(\langle\varphi^4\rangle_0\langle
\varphi^2\rangle_0-\langle \varphi^6\rangle_0\right)
+\varepsilon_{K}\frac{3}{2}\left(\frac{\sigma^2}{K}\right)^3=0 .
\end{equation}
Again, the only acceptable solution to this equation is $\varepsilon_K=0$.
Therefore, if there exists a critical point in the phase
diagram where a continuous disorder-multistability
phase transition occurs, that point is singular in the sense that
it is isolated.


\begin{thebibliography}{99}
\bibitem{nota1}  Note that the present analysis is also applicable to
equilibrium models with $\Gamma =1$, which leads to 
to the so-called Gaussian model when $V=V_{1}$ and to model A
if the local potential is $V_{2}$.
\end{thebibliography}
\end{document}